\documentclass[fleqn,usenatbib]{mnras}

\usepackage{newtxtext,newtxmath}
\usepackage[T1]{fontenc}

\DeclareRobustCommand{\VAN}[3]{#2}
\let\VANthebibliography\thebibliography
\def\thebibliography{\DeclareRobustCommand{\VAN}[3]{##3}\VANthebibliography}

\usepackage{graphicx}	% Including figure files
\usepackage{amsmath}	% Advanced maths commands
\usepackage{amssymb}	% Extra maths symbols

\title[Spectro-temporal structure of FRB 121102]{A simple relationship for the spectro-temporal structure of bursts from FRB 121102}

\author[F. Rajabi et al.]{Fereshteh Rajabi,$^{1,2}$\thanks{E-mail: f3rajabi@uwaterloo.ca}
Mohammed A. Chamma,$^{3}$ Christopher M. Wyenberg,$^{3}$ 
\newauthor Abhilash Mathews$^{4}$ and Martin Houde$^{3}$\thanks{E-mail: mhoude2@uwo.ca}
\\
$^{1}$Perimeter Institute for Theoretical Physics, Waterloo, ON N2L 2Y5, Canada\\
$^{2}$Institute for Quantum Computing and Department of Physics and Astronomy, The University of Waterloo, 200 University Ave. West, \\Waterloo, Ontario N2L 3G1, Canada\\
$^{3}$Department of Physics and Astronomy, The University of Western Ontario, 1151 Richmond Street, London, Ontario N6A 3K7, Canada\\
$^{4}$Plasma Science and Fusion Center, Massachusetts Institute of Technology, 77 Massachusetts Avenue, Cambridge, MA 02139, USA\\}

\date{Accepted 2020 September 3. Received 2020 August 30; in original form 2020 June 25}

\pubyear{2015}

\begin{document}
\label{firstpage}
\pagerange{\pageref{firstpage}--\pageref{lastpage}}
\maketitle

\begin{abstract}
We consider a simple dynamical and relativistic model to explain the spectro-temporal structure often displayed by repeating fast radio bursts (FRBs). We show how this model can account for the downward frequency drift in a sequence of sub-bursts of increasing arrival time (the ``sad trombone'' effect) and their tendency for exhibiting a reduced pulse width with increasing frequency of observation. Most importantly, this model also predicts a systematic inverse relationship between the (steeper) slope of the frequency drift observed within a single sub-burst and its temporal duration. Using already published data for FRB 121102 we find and verify the relationship predicted by this model. We therefore argue that the overall behaviour observed for this object as a function of frequency is consistent with an underlying narrow-band emission process, where the wide-band nature of the measured FRB spectrum is due to relativistic motions. Although this scenario and the simple dynamics we consider could be applied to other theories, they are well-suited for a model based upon Dicke's superradiance as the physical process responsible for FRB radiation in this and similar sources.               
\end{abstract}


%%%%%%%%%%%%%%%%%%%%%%%%%%%%%%%%%%%%%%%%%%%%%%%%%%

%%%%%%%%%%%%%%%%% BODY OF PAPER %%%%%%%%%%%%%%%%%%

\section{Introduction}\label{sec:introduction}

%1. A short introduction on repeating FRBs and the sad trombone feature.\\
Since the discovery of the first fast radio burst (FRB; \citealt{Lorimer2007}), a large number of FRBs have been detected using different telescopes. Among these FRB 121102 consisted of the first and only repeating FRB reported in over a decade up until 2018, when a second repeater (FRB 180814.J422+73) was discovered with the Canadian Hydrogen Intensity Mapping Experiment (CHIME) facility \citep{CHIME2019a}. This discovery, and others to follow \citep{CHIME2019b,CHIME2020}, not only ruled out the uniqueness of FRB 121102 but further suggested that a large population of repeaters could be found using highly sensitive observational experiments. 

Accordingly, at the time of writing the number of reported repeaters is close to 20 \citep{CHIME2020} while the total population of FRBs, including both one-off events and repeaters, stands at around 700. These numbers are growing rapidly as experiments such as CHIME/FRB sample the sky daily. While there are several theoretical proposals, questions concerning the origin of FRBs remain unsolved. However, as more data become available, especially as the number of repeaters increases, more FRBs can be localized to their host environments \citep{Day2020} and shed light on the physical mechanism producing these high-energy millisecond pulses. 

The monitoring of repeaters has also provided important information on the spectro-temporal characteristics of these bursts. For example, an interesting feature reported for repeating FRB signals is a downward drift in the frequency of sub-bursts with increasing arrival time. This is the so-called ``sad trombone'' effect. The two-dimensional auto-correlation analyses of dynamic spectra of FRB 121102 by \citet{Hessels2019} and \citet{CHIME2019c} quantified this downward drift in some of the bursts obtained at different frequency bands for this source. Although not all repeating FRBs show the sad trombone behaviour, the second discovered repeater, FRB 180814.J0422+73, and several other such sources also exhibit this effect (e.g., sources 1, 3, 4, 5, 6, 7 and 8 reported in \citealt{CHIME2019b}). 

In this paper, we present a simple dynamical and relativistic model aimed at explaining the details of the spectro-temporal structure of FRB 121102 and other similar sources. In particular, we show how observed characteristics pertaining to the temporal width of sub-bursts and the aforementioned sad trombone effect are inherent to the model. Most importantly, we also predict a systematic inverse relationship between the frequency drift within a single sub-burst and its temporal duration, which we verify using existing data. 

The presentation is structured as follows. We discuss our model in Section \ref{sec:model}, while focusing on the aforementioned predictions and their verification using material found in the existing literature in Section \ref{sec:predictions}. As previously mentioned, we concentrate on FRB 121102 because of the wealth of data available on this source. Finally, we discuss the implications of our model for the nature of the physical process underlying FRB emission in this source, and its consistency with Dicke's superradiance \citep{Houde2018a,Houde2018b}. 

\section{Model and observational evidence}\label{sec:model}
\begin{center}
    \begin{figure}
        \centering
        \includegraphics[trim=0 120 0 40 , clip, width=0.9\columnwidth]{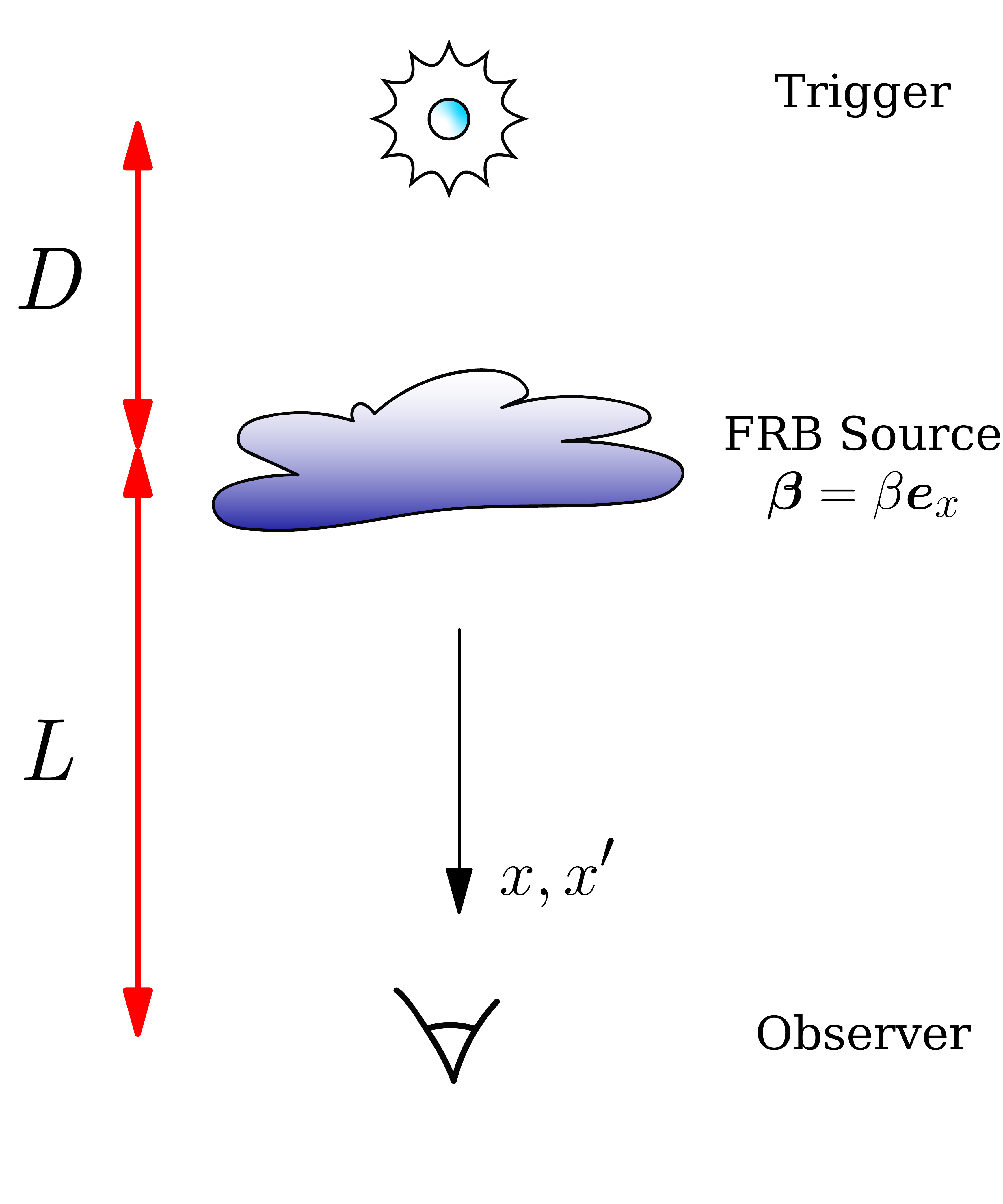}
        \caption{Schematic of the system considered for our FRB model. It consists of a trigger source, an FRB source and an observer. All are assumed to be aligned along the line of sight from the observer, with the trigger source located behind the FRB. The FRB source is moving at a relativistic velocity $\mathbf{v}=\boldsymbol{\beta}c=\beta c\mathbf{e}_x$ relative to the observer.}
        \label{fig:cartoon}
    \end{figure}
\end{center}

The physical model we are considering for our analysis is presented in Figure \ref{fig:cartoon}. The system consists of a trigger source, an FRB source, and an observer. These components are assumed to be aligned along a single axis with the trigger located behind the FRB source, as seen by the observer. The FRB source is moving at a relativistic velocity $\mathbf{v}=\boldsymbol{\beta}c=\beta c\mathbf{e}_x$ relative to the observer, where $\beta$ can be positive or negative ($c$ is the speed of light). We thus define two reference frames: the rest frame of the FRB, with axes $t^\prime$ and $x^\prime$, and the observer's reference frame, with axes $t$ and $x$ (note that $x^\prime$ and $x$ are aligned). The space-time coordinates of an event measured in the observer's frame are related to those measured in the rest frame of the FRB through the Lorentz transformation
\begin{align}
    & t^\prime = \gamma\left(t-\frac{\beta}{c}x\right)\label{eq:t'}\\
    & x^\prime = \gamma\left(x-\beta ct\right)\label{eq:x'}
\end{align}
\noindent with $\gamma=1/\sqrt{1-\beta^2}$.\\

\begin{center}
    \begin{figure}
        \centering
        \hspace*{-0.7cm}
        \includegraphics[trim=100 50 115 40, clip, width=1\columnwidth]{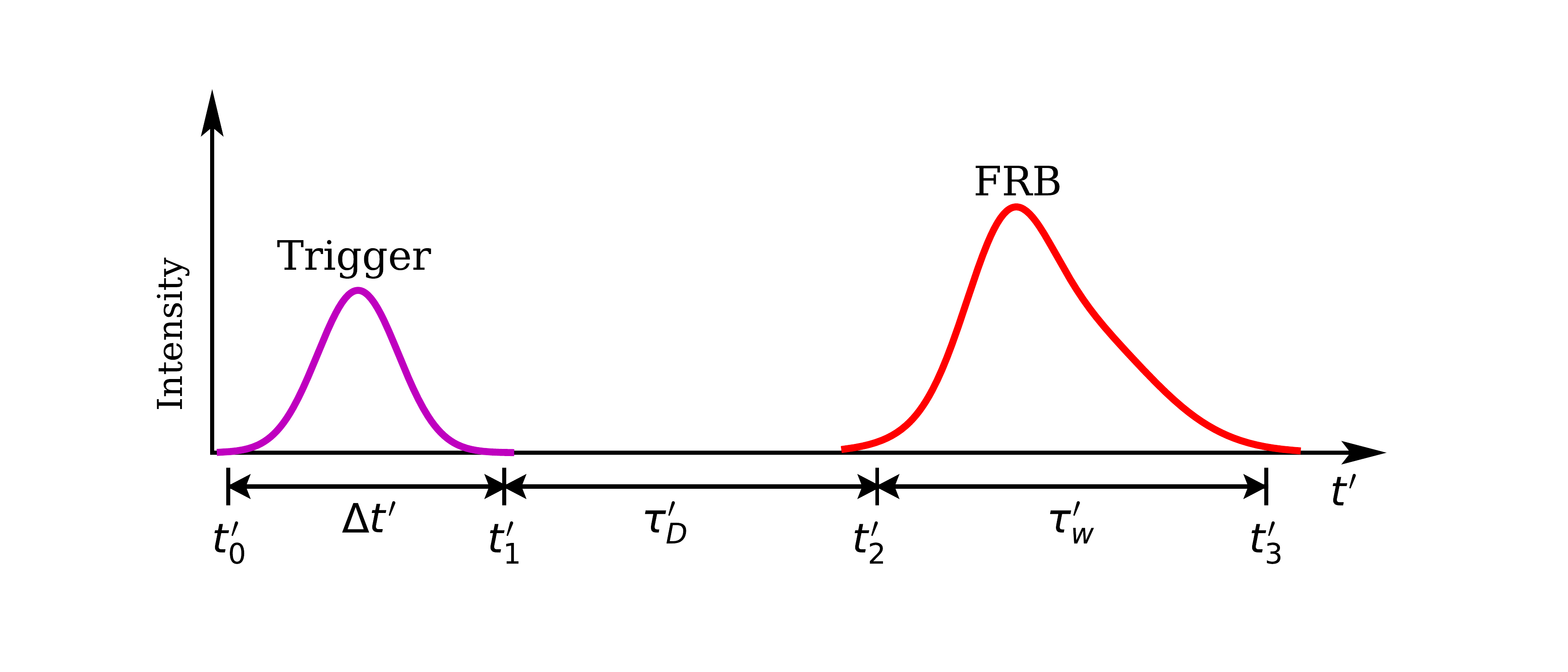}
        \caption{Time sequence for the trigger signal, the delay time $\tau^\prime_\mathrm{D}$ and the burst duration $\tau^\prime_\mathrm{w}$ as seen in the reference frame of the FRB source (not to scale).}
        \label{fig:time_sequence}
    \end{figure}
\end{center}

The arrival of a signal from the trigger at the FRB source stimulates and is followed by, after some delay, the emergence of an intensity burst. The time sequence of the trigger signal and the FRB pulse as viewed in the FRB reference frame is shown in Figure \ref{fig:time_sequence}. Still in the FRB frame, the proper time-scale $\tau^\prime_\mathrm{D}$ stands for the aforementioned time delay before the emergence of the intensity burst after the arrival of the trigger signal, while the temporal duration of the FRB signal is denoted by $\tau^\prime_\mathrm{w}$.

Using equations (\ref{eq:t'})-(\ref{eq:x'}) and accounting for the distance travelled by the FRB source relative to the observer in corresponding time intervals, the time delay and duration associated to the FRB pulse are measured by the observer to be
\begin{align}
    & t_\mathrm{D} = \tau^\prime_\mathrm{D}\frac{\nu_0}{\nu_\mathrm{obs}}\label{eq:t_D}\\
    & t_\mathrm{w} = \tau^\prime_\mathrm{w}\frac{\nu_0}{\nu_\mathrm{obs}}.\label{eq:t_w}
\end{align}
\noindent In these equations the frequency of the radiation in the FRB rest frame $\nu_0$ and that detected in the observer's frame $\nu_\mathrm{obs}$  are related through the relativistic Doppler shift formula
\begin{equation}
    \nu_\mathrm{obs} = \nu_0\sqrt{\frac{1+\beta}{1-\beta}}.\label{eq:Doppler}
\end{equation}

\begin{center}
    \begin{figure}
        \centering
        \hspace*{-0.1cm}
        \includegraphics[trim=20 40 20 50, clip, width=1.\columnwidth]{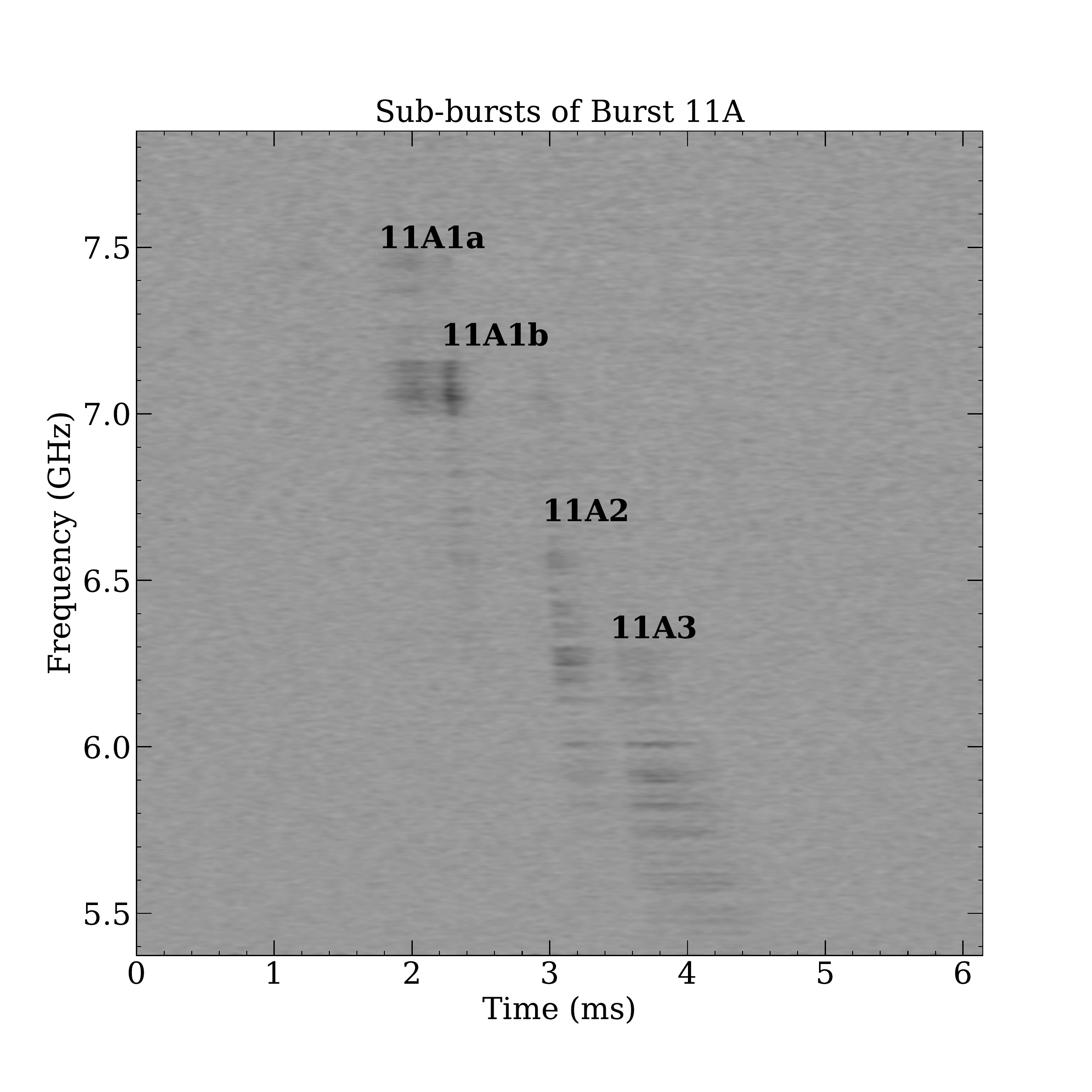}
        \caption{Burst 11A detected in FRB 121102 by \citet{Gajjar2018}. The different sub-bursts are identified and their relative drift is clearly seen. Many other such examples exist in the literature at different frequency bands for FRB 121102 (e.g., \citealt{Hessels2019}) and other repeaters \citep{CHIME2019b,CHIME2020}.} 
        \label{fig:Gajjar}
    \end{figure}
\end{center}

Before discussing the consequences ensuing from this model, it will be beneficial to define the terminology used to describe the spectro-temporal structure of FRBs. To do so we show in Figure \ref{fig:Gajjar} the example of Burst 11A detected in FRB 121102 by \citet{Gajjar2018}. As can be seen, four separate intensity pulses, labelled 11A1a, 11A1b, 11A2, and 11A3 in the figure, are identified. Following the literature, we will refer to these intensity features as ``sub-bursts.'' Likewise, there has been ample discussions of frequency drifts with arrival time in the signals of repeating FRBs \citep{Gajjar2018,Hessels2019,CHIME2019c,CHIME2019b}. Here, we differentiate between two kinds of frequency drifts: \textit{i)} the relative downward drift in the frequency of sub-bursts with increasing arrival time, generally referred to as the sad trombone effect, and \textit{ii)} the steeper downward frequency drift within a single sub-burst; we define this behaviour as the ``sub-burst drift.'' Both frequency drifts can be explained using our simple model.

In general there can be more than one FRB source located along the line-of-sight between the observer and the trigger source. It is important to note that equations (\ref{eq:t'}) to (\ref{eq:Doppler}) apply equally well to all these sources, even in cases where they are located at different distances from the observer. That is, if two such FRB sources emit identical sub-bursts, i.e., at the same frequency $\nu_0$ and with the same time-scales $\tau^\prime_\mathrm{D}$ and $\tau^\prime_\mathrm{w}$ in their local rest frames, their signals would be indistinguishable for the observer (i.e., they would be detected at the same time and would have the same duration) whenever the frames move at the same velocity relative to the observer. This is because the more distant FRB sources along the line of sight will burst earlier than sources that are closer. Within the context of our model, measured differences in time of arrival and duration between identical sub-bursts in the FRB source are solely due to differences in their respective velocity $\beta$ relative to the observer.

More generally, the difference in arrival time between any sub-bursts emitted at the same frequency $\nu_0$ in their individual rest frames is calculated, using equation (\ref{eq:t_D}), to be
\begin{equation}
    \Delta t_\mathrm{D} = -t_\mathrm{D}\left(\frac{\Delta\nu_\mathrm{obs}}{\nu_\mathrm{obs}}-\frac{\Delta\tau^\prime_\mathrm{D}}{\tau^\prime_\mathrm{D}}\right),\label{eq:Dt_D}
\end{equation}
\noindent where $\Delta\nu_\mathrm{obs}$ is the change in the observed frequency due to a velocity spread within a source responsible for a single sub-burst while $\Delta\tau^\prime_\mathrm{D}$ accounts for variations in proper delay time among the sub-bursts.

\subsection{Predictions made by the model -- FRB 121102}\label{sec:predictions}

Equations (\ref{eq:t_D}) to (\ref{eq:Dt_D}), although very simple, have profound implications. In what follows we discuss some predictions that can be drawn from them concerning the characteristics of FRB signals, as detected in the observer's reference frame.

\subsubsection{Sub-burst duration vs. frequency of observation}

A readily measurable parameter of FRB pulses is their time duration. To be precise, in cases where multiple bursts are detected during one event, as in Figure \ref{fig:Gajjar}, we focus on the sub-burst width $t_\mathrm{w}$ as opposed to the duration of the whole event. Equation (\ref{eq:t_w}) clearly shows that for a given proper time-scale $\tau^\prime_\mathrm{w}$ and emission frequency $\nu_0$ in the FRB rest frame one should expect shorter sub-bursts with increasing frequency of observation $\nu_\mathrm{obs}$. A similar behaviour is predicted for the observed time delay $t_\mathrm{D}$ (see equation \ref{eq:t_D}). However, since the time delay cannot be unambiguously determined for a given event (i.e., the time of the trigger is unknown), it is advantageous to focus on  $t_\mathrm{w}$ instead of $t_\mathrm{D}$.

Although this behaviour is visually apparent in individual sub-bursts (see Figure \ref{fig:Gajjar}), the effect should be more pronounced between observations made in different frequency bands. However, as will be seen in Sec. \ref{sec:sub-burst} when using the FRB 121102 data of \citet{Michilli2018}, observations at a given frequency reveal a significant spread in sub-burst durations (i.e., $t_\mathrm{w}$ covers a wide range of values). It follows that experimental verification of the prediction of a decrease in burst duration with increasing frequency requires a significant amount of data over several frequency bandwidths. Although this complicates such a test, a decrease in burst duration with increasing frequency has already been noted using comparisons of observations obtained in different frequency bands for FRB 121102 \citep{Gajjar2018,Hessels2019}. In particular, \citealt{Gajjar2018} summarize some of the past observational data on FRB 121102 and clearly show the narrowing of signal at higher frequencies in their Figure 7b\footnote{Although the decrease in burst duration with increasing $\nu_\mathrm{obs}$ is clearly visible in Figure 7b of \citet{Gajjar2018}, the effect is likely more pronounced than seen there as the authors do not measure the sub-burst duration $t_\mathrm{w}$ but rather the duration of the whole event. This provides an upper limit for $t_\mathrm{w}$, especially at higher frequencies where multiple sub-bursts are more likely to be detected (as in Figure \ref{fig:Gajjar}).}. 

\subsubsection{Sub-burst drift vs. sub-burst duration}\label{sec:sub-burst}

We now discuss a striking feature that can be extracted from the dynamic spectra of FRB signals from repeaters, either from lone bursts or individual sub-bursts. More precisely, we focus on the variation in frequency with time within a given sub-burst, which, as mentioned earlier, we refer to as sub-burst drift. This feature is not to be confused with the sad trombone effect to be discussed later, which concerns the relative drift of the central frequency in a sequence of sub-bursts.  The slope of the sub-burst drift is more pronounced than that observed for the sad trombone (see Figure \ref{fig:Gajjar}). 

\begin{center}
    \begin{figure}
        \centering
        % \hspace*{-0.3cm}
        % \includegraphics[trim=0 120 65 180, clip, width=1.\columnwidth]{burst2and3.pdf}
        %%% Use this for bigger, will be on the next page however
        %\hspace*{-0.6cm}
        \includegraphics[trim=40 120 60 180, clip, width=1.\columnwidth]{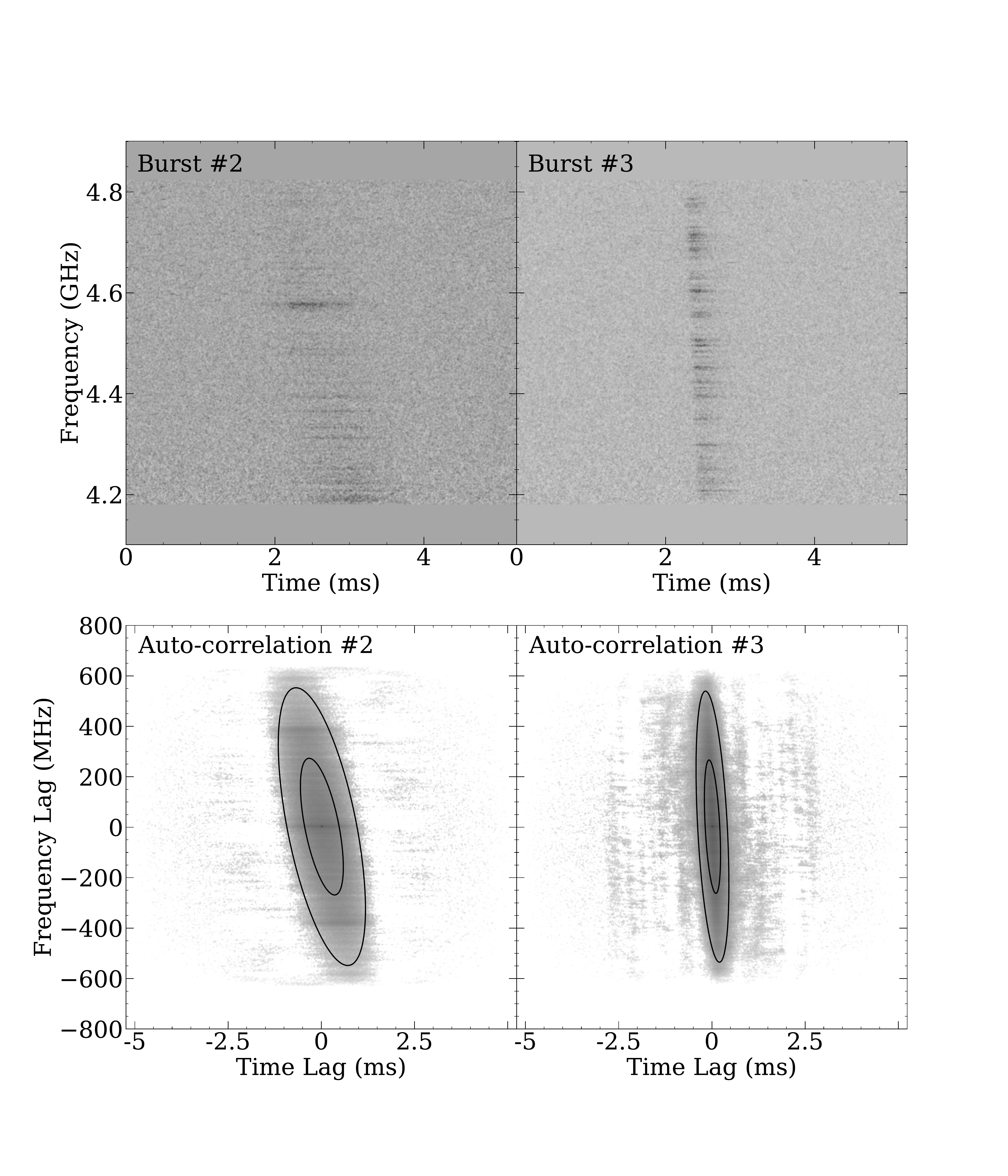}
        \caption{\textit{(Top)} Dynamic spectra of Bursts \#2 and \#3 from \citet{Michilli2018} with \textit{(bottom)} their two-dimensional autocorrelation functions \citep{Hessels2019}. The contours plotted on each autocorrelation function are for the one- and two-standard deviation levels of a two-dimensional Gaussian fit. The corresponding sub-burst drifts for \#2 and \#3 are $-741$~MHz/ms and $-2824$~MHz/ms, respectively. Note how the steeper slope for Burst \#3 corresponds to a shorter burst duration.}
        \label{fig:Michilli}
    \end{figure}
\end{center}

We define the sub-burst drift as the corresponding slope $d\nu_\mathrm{obs}/dt_\mathrm{D}$ observed in the dynamic spectrum of a signal. Within the context of our model, we can find an expression for this quantity by setting $\Delta\tau^\prime_\mathrm{D}=0$ in equation (\ref{eq:Dt_D}) (i.e., focusing on a single sub-burst) and thus obtain
\begin{align}
    \frac{d\nu_\mathrm{obs}}{dt_\mathrm{D}} & = -\frac{\nu_\mathrm{obs}}{t_\mathrm{D}}\nonumber\\
    & = -\left(\frac{\tau^\prime_\mathrm{w}}{\tau^\prime_\mathrm{D}}\right)\frac{\nu_\mathrm{obs}}{t_\mathrm{w}}, \label{eq:sub_drift}
\end{align}
\noindent for the slope at the beginning of the sub-burst. The ratio of equations (\ref{eq:t_D}) and (\ref{eq:t_w}) was used for the last relation. This was done because $t_\mathrm{w}$ is readily measurable from the data while, as mentioned earlier, the delay time $t_\mathrm{D}$ is not known in the absence of the trigger signal. 

Equation (\ref{eq:sub_drift}) predicts that for a given frequency $\nu_\mathrm{obs}$, the sub-burst drift scales inversely with its time duration $t_\mathrm{w}$. To verify this behaviour, we have extracted $d\nu_\mathrm{obs}/dt_\mathrm{D}$ and $t_\mathrm{w}$ from some data available for FRB 121102. More precisely, both quantities were quantified using the autocorrelation function technique introduced by \citet{Hessels2019} on the dynamic spectra presented in \citet{Michilli2018}, \citet{Gajjar2018} (only Bursts 11A (and its sub-bursts) and 11D) and \citet{CHIME2019c} (one CHIME/FRB detection centred at $\sim630$~MHz). We show two examples taken from the \cite{Michilli2018} data in Figure \ref{fig:Michilli}. In all cases the autocorrelation (bottom panels) of the dynamic spectrum (top panels) was fitted to a two-dimensional Gaussian function, where the orientation of the minor and major axes relative to the vertical and horizontal axes reveal the sub-burst drift $d\nu_\mathrm{obs}/dt_\mathrm{D}$ while the intersection of the one-standard deviation contour with the horizontal axis provides us with an estimate of the duration $t_\mathrm{w}$. The behaviour predicted by equation (\ref{eq:sub_drift}) can be clearly seen in these examples, i.e., the shorter pulse having a steeper slope $d\nu_\mathrm{obs}/dt_\mathrm{D}$. 

\begin{center}
    \begin{figure*}
        \centering
        \includegraphics[trim=55 25 80 65, clip,width=\textwidth]{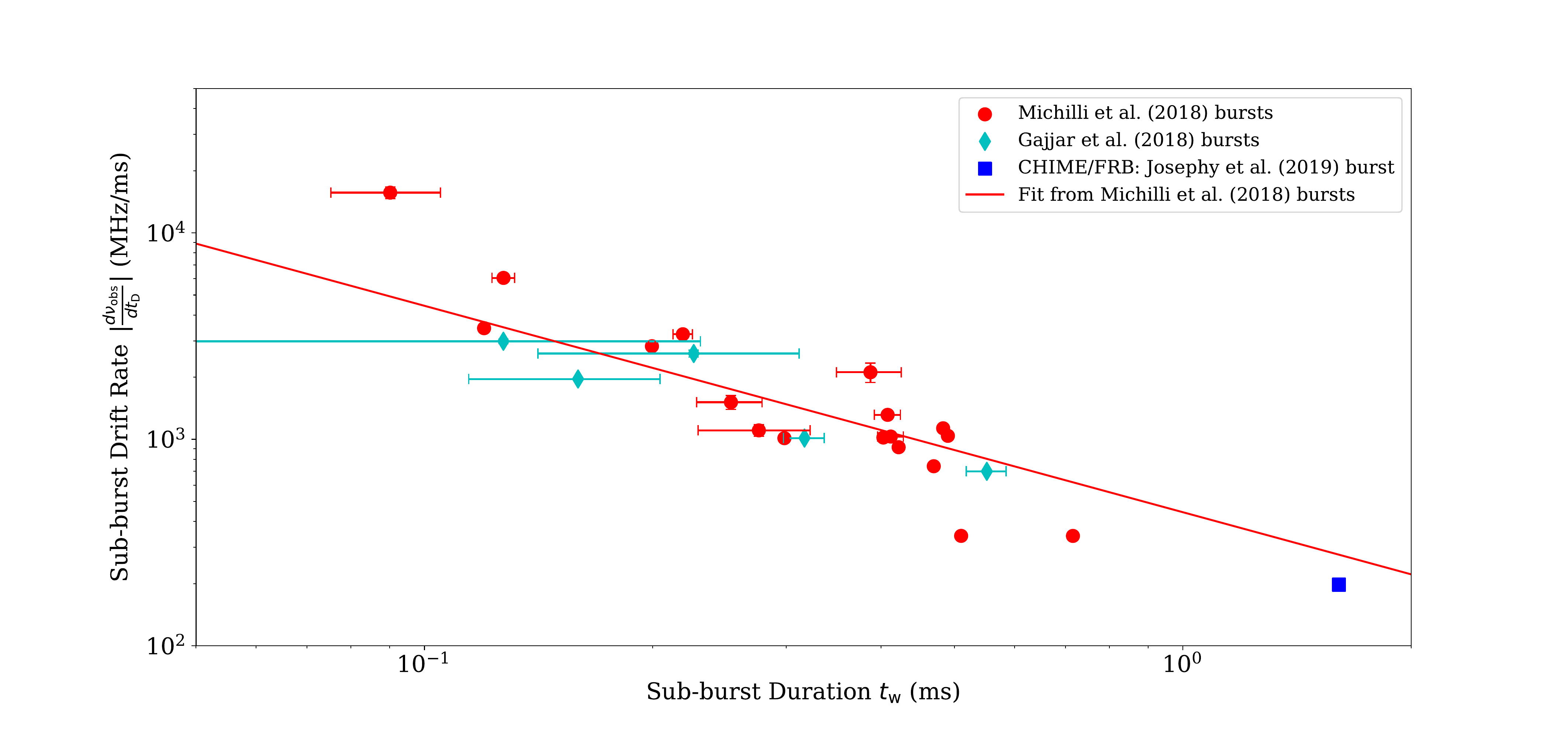}
        \caption{A plot of $\left|d\nu_\mathrm{obs}/dt_\mathrm{D}\right|$ vs. $t_\mathrm{w}$ on logarithmic scales for the data of \citet{Michilli2018} (red circles), \citet{Gajjar2018} (teal diamonds; only Bursts 11A (and its sub-bursts) and 11D) and \citet{CHIME2019c} (blue square; one datum). These parameters were extracted using the two-dimensional autocorrelation technique of \citet{Hessels2019}.The solid red curve is for the fit $\left|d\nu_\mathrm{obs}/dt_\mathrm{D}\right|=A \nu_\mathrm{obs}/t_\mathrm{w}$ on the \citet{Michilli2018} data (i.e., $\nu_\mathrm{obs}=4.47$~GHz); $A\simeq0.1$. The \citet{Gajjar2018} and \citet{CHIME2019c} data were scaled according to their frequency of observations relative to the \citet{Michilli2018} data (see text). Error bars shown only account for uncertainties derived from the underlying Gaussian fit and do not reflect uncertainties caused by, for example, de-dispersion of the data.} 
        \label{fig:sub-burst_drift}
    \end{figure*}
\end{center}

Our results are summarized in Figure \ref{fig:sub-burst_drift}, where we plot $\left|d\nu_\mathrm{obs}/dt_\mathrm{D}\right|$ vs. $t_\mathrm{w}$ on logarithmic scales for all the data published in \citet{Michilli2018} (red circles), as well as that from \citet{Gajjar2018} (teal diamonds) and \citet{CHIME2019c} (blue square, one datum). Whenever possible we identified and separated sub-bursts within a dynamic spectrum (as shown in Figure \ref{fig:Gajjar} for Burst 11A of \citealt{Gajjar2018}) and ran the two-dimensional autocorrelation analysis separately on each feature. Although it was at times difficult to determine when and where should such a procedure be applied, the systematic temporal narrowing of sub-bursts with steeper drifts is clearly seen in the data. The solid red curve in the figure is the result of a fit of the type $\left|d\nu_\mathrm{obs}/dt_\mathrm{D}\right|=A\nu_\mathrm{obs}/t_\mathrm{w}$ to the \citet{Michilli2018} data alone (i.e., $\nu_\mathrm{obs}=4.47$~GHz, see below), where we find $A=\tau^\prime_\mathrm{w}/\tau^\prime_\mathrm{D}\simeq0.1$.
 
It should be noted that since the relation between $d\nu_\mathrm{obs}/dt_\mathrm{D}$ and $t_\mathrm{w}$ also depends on $\nu_\mathrm{obs}$ (see equations \ref{eq:t_D} and \ref{eq:sub_drift}), one needs to scale data points obtained at different frequencies in order to meaningfully compare them. More precisely, insertion of equation (\ref{eq:t_w}) in equation (\ref{eq:sub_drift}) shows that $d\nu_\mathrm{obs}/dt_\mathrm{D}\propto\nu_\mathrm{obs}^2/\nu_0$, while $t_\mathrm{w}\propto\nu_0/\nu_\mathrm{obs}$. If we assume that data obtained in different frequency bands (e.g., $\sim4.5$~GHz for \citealt{Michilli2018}, $\sim6$--$8$~GHz for \citealt{Gajjar2018} and $\sim600$--$700$~MHz for \citealt{CHIME2019c}) were emitted at the same frequency $\nu_0$ in their respective rest frames, then we must choose one data set as a reference (we selected the \citealt{Michilli2018} data set for this) and calibrate the others against it for direct comparison. For example, the sub-burst drifts for the \citet{Gajjar2018} data were scaled down through a multiplication by $\left(\nu_\mathrm{M}/\nu_\mathrm{G}\right)^2$ and their duration scaled up by $\nu_\mathrm{G}/\nu_\mathrm{M}$, with $\nu_\mathrm{M}$ and $\nu_\mathrm{G}$ the centre frequencies for the \citet{Michilli2018} and \citet{Gajjar2018} sub-bursts, respectively. We adopted a single frequency $\nu_\mathrm{M}=4.47$~GHz for all the \citet{Michilli2018} sub-bursts, while we adjusted $\nu_\mathrm{G}$ for those from \citet{Gajjar2018} to match the observed frequencies. The same process was applied for the CHIME/FRB datum centred at $\sim630$~MHz \citep{CHIME2019c}. The fact that three sets of data, spanning more than a decade in frequency, appear to follow the same law for their respective sub-burst drifts in Figure \ref{fig:sub-burst_drift} is significant. As we will discuss in Sec. \ref{sec:discussion}, it is strong evidence in favour of having a single rest frame frequency of emission $\nu_0$ for all bursts detected regardless of the frequency of observation. 

\subsubsection{Relative drift between sub-bursts -- the sad trombone}

In cases when multiple sub-bursts are present in signals from repeating FRBs, it is often observed that the characteristic frequency tends to drift downward with increasing arrival time. This relative frequency drift between sub-bursts, i.e., the sad trombone effect, is contained in equation (\ref{eq:Dt_D}) of our model. More precisely, it is accounted for through the second term on the right-hand side of this equation, which describes the expected difference in arrival time for sub-bursts of varying proper time delay $\tau^\prime_\mathrm{D}$. Focusing on this term only, and allowing for a dependency in the proper delay time with the local velocity within the FRB source (and thus with the observed frequency $\nu_\mathrm{obs}$) we write $\Delta\tau^\prime_\mathrm{D}=\Delta\nu_\mathrm{obs}\,d\tau^\prime_\mathrm{D}/d\nu_\mathrm{obs}$ and find
\begin{equation}
    \frac{\Delta\nu_\mathrm{obs}}{\Delta t_\mathrm{D}} = \frac{\nu_\mathrm{obs}}{\nu_0}\frac{d\nu_\mathrm{obs}}{d\tau^\prime_\mathrm{D}},\label{eq:relative_drift}
\end{equation}
where we also used equation (\ref{eq:t_D}). The ``$\Delta$'' notation is used here to distinguish this effect from the sub-burst drift discussed in Sec. \ref{sec:sub-burst}.

Equation (\ref{eq:relative_drift}) shows that the sign of the relative frequency drift depends on the parameter $d\nu_\mathrm{obs}/d\tau^\prime_\mathrm{D}$. Whenever $d\nu_\mathrm{obs}/d\tau^\prime_\mathrm{D}<0$ the relative drift will be descending (i.e., of negative slope) with the arrival of sub-bursts, leading to the appearance of the sad trombone effect in the corresponding dynamic spectrum. As sub-bursts arrive over time, their central frequency drifts to lower values. This behaviour has already been observed for FRB 121102 and other repeating FRBs \citep{CHIME2019b,CHIME2020}. Salient examples of this feature can be clearly seen for Burst 11A of Gajjar et al. (2018) shown in Figure \ref{fig:Gajjar}, as well as in other observations of FRB 121102 \citep{Hessels2019}.

On the other hand, whenever $d\nu_\mathrm{obs}/d\tau^\prime_\mathrm{D}>0$ the opposite behaviour is expected (i.e., a ``happy trombone'' feature). It is interesting to note that such a behaviour has recently been observed in the FRB-like emission detected in the direction of the Galactic magnetar SGR~1935+2154 \citep{Scholz2020,CHIME2020b} and also for FRB 190611 \citep{Day2020}.

Most importantly, equation (\ref{eq:relative_drift}) also predicts that the relative frequency drift should be more pronounced with increasing frequency of observation $\nu_\mathrm{obs}$, for a fixed rest frame frequency of emission $\nu_0$. This has already been verified for FRB 121102 by \citet{Hessels2019} through comparison of the relative sub-burst drift for measurements obtained from approximately $1$--$8$~GHz; see their Figure 3 (top left and right panels). For example, their analysis (Fig. 3, right panel) indicates a frequency drift that is approximately four times stronger at $\sim6.5$~GHz than at $\sim2$~GHz (i.e., $\sim-800$~MHz/ms vs. $\sim-200$~MHz/ms). This is consistent with the behaviour predicted by equation (\ref{eq:relative_drift}) under the assumption of a common rest frame frequency of emission $\nu_0$ and parameter $d\nu_\mathrm{obs}/d\tau^\prime_\mathrm{D}$ for all observations.   

\section{Discussion and summary}\label{sec:discussion}

As we have seen in the previous section, our dynamical model, despite its simplicity, makes predictions that are verified using data available for FRB 121102. Perhaps most strikingly, it reveals a close relationship between the sub-burst drift and duration. For FRB 121102 this last prediction is corroborated with the results presented in Figure \ref{fig:sub-burst_drift}. Significantly, we once again emphasize the fact that the fit to the data based on equation (\ref{eq:sub_drift}) was performed on those from \citet{Michilli2018} alone while the other data from \citet{Gajjar2018} and \citet{CHIME2019c} were found to agree remarkably well with it. As we mentioned earlier, the fact that data sets covering more than a decade in frequency behave in a consistent manner is an important clue to the nature of the physical process underlying FRB signals in this source. Indeed, this agreement makes it difficult to imagine how these data could have resulted from emission taking place at different frequencies in their corresponding rest frames. This statement rests on the relationship between the sub-burst drift $d\nu_\mathrm{obs}/dt_\mathrm{D}$ and duration $t_\mathrm{w}$ through the frequency of emission $\nu_0$ in the FRB rest frame, as displayed in equation (\ref{eq:sub_drift}). In other words, it would not be possible for the three data sets to share the same law (i.e., the fit in Figure \ref{fig:sub-burst_drift}) linking these two parameters if they did not also share the same frequency $\nu_0$.

As far as we can tell from available data, practically all sub-bursts from repeating FRBs can have an associated frequency drift. For the FRB 121102 data used for our analysis the extent of the frequency drift can range from 0.3~GHz to 2~GHz, but it is difficult to be specific because sometimes the (sub-)burst extends beyond the observational frequency bandwidth. Once a dispersion measure (DM) is chosen and as long as all the bursts are de-dispersed to the same DM, the existence of a relationship between the sub-burst drift and duration is independent of the DM. Although applying a de-dispersion with a greater or lesser DM will certainly change the drift accordingly within a set of sub-bursts, the relative importance of the drift from one sub-burst to the other is preserved. For bursts from non-repeaters the prevalence of sub-burst drifts is difficult to assess because the de-dispersion is purposefully done so that the frequency drift (proportional to $\nu_\mathrm{obs}^{-2}$) is removed (there are no sub-burst trains from non-repeaters, so it is not possible to implement a de-dispersion that optimizes the structure of the burst as is often done for repeaters).

The earlier comment concerning a unique frequency $\nu_0$ also applies for the sad trombone effect common to several FRBs, including FRB 121102 (see Figure \ref{fig:Gajjar}). As mentioned earlier, the measured frequency drift between sub-bursts scales approximately linearly with the observed frequency $\nu_{\mathrm{obs}}$ (see top panels in Figure 3 of \citet{Hessels2019}). Based on equation (\ref{eq:relative_drift}), our model predicts the slope of the relative drift $\Delta\nu_\mathrm{obs}/\Delta t_\mathrm{D}$ with observed frequency $\nu_\mathrm{obs}$ to vary linearly with the parameter $d\nu_\mathrm{obs}/d\tau^\prime_\mathrm{D}$ and inversely with the rest frame frequency of emission $\nu_0$. It follows that, similarly to the case of sub-burst frequency drift discussed above, the observed behaviour is most easily explained if we are in the presence of a narrow-band emission process at frequency $\nu_0$ in all FRB rest frames. The functionality of the two effects, i.e., the sub-burst drift and the sad trombone, with the observed frequency $\nu_{\mathrm{obs}}$ provides strong evidence in favour of such a scenario. 
\begin{center}
    \begin{figure}
        \centering
        %\hspace*{-0.2cm}
        \includegraphics[trim=0 0 25 20, clip,width=1.\columnwidth]{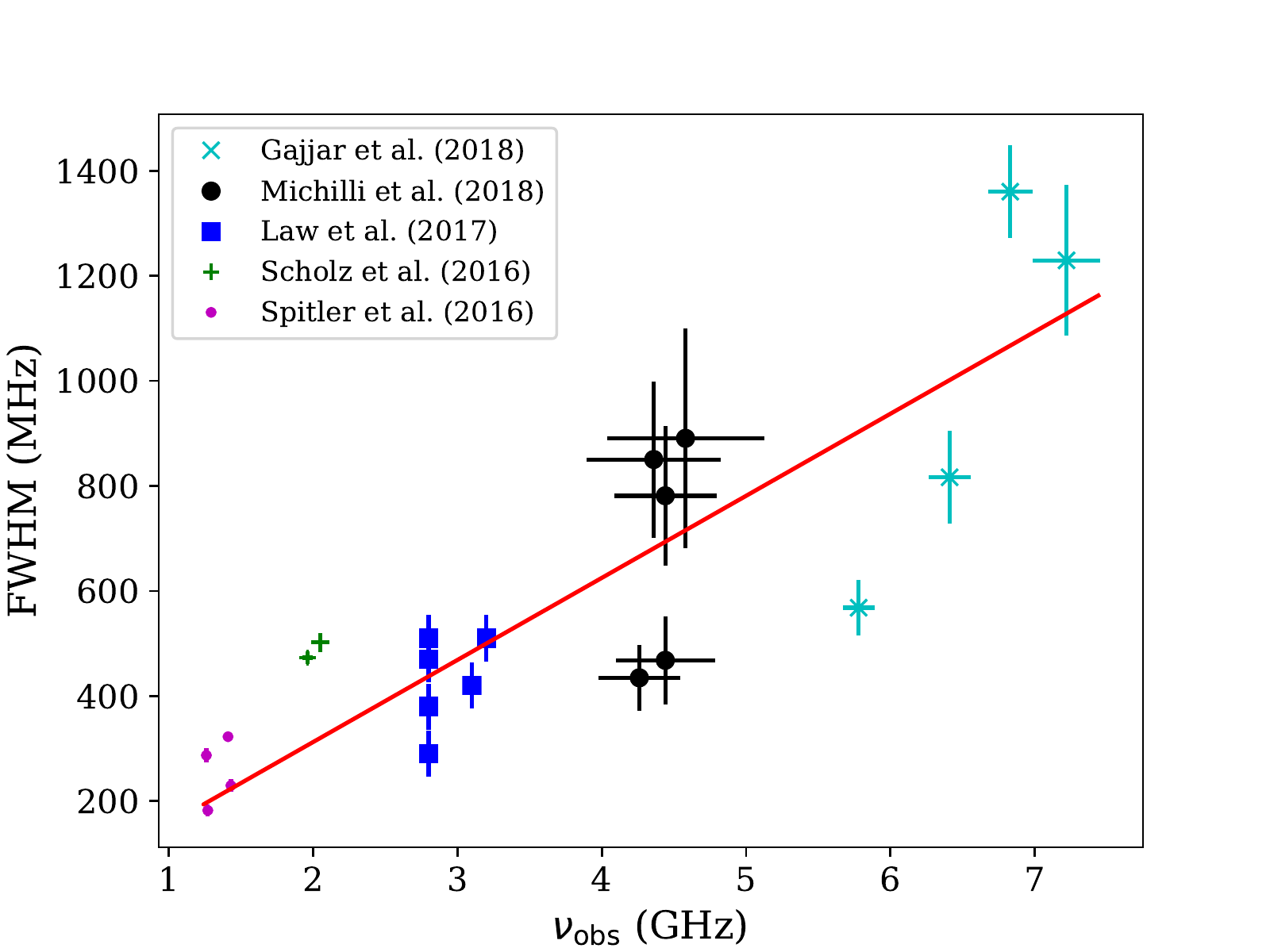}
        \caption{Evidence for relativistic motions in FRB 121102, taken from \citet{Houde2018b}.  The bandwidth (FWHM in MHz) of FRB 121102 sub-bursts measured as a function of the frequency band at which they were measured. The solid red line is a linear fit yielding a slope $156\pm4$~MHz/GHz, implying relativistic motions (see text).
} 
        \label{fig:Houde_fig}
    \end{figure}
\end{center}

If FRB 121102 and similar sources host an underlying narrow-band emission process, then relativistic motions are needed to explain the frequency extent, spanning several GHz, of the detected signals. More precisely, in such cases the frequency measured by the observer will not scale linearly with the velocity of the source but rather follow the relativistic Doppler shift formula given in equation (\ref{eq:Doppler}). Indeed, evidence for relativistic motion in FRB 121102 already exists. In Figure \ref{fig:Houde_fig} we show the results previously published by \citet{Houde2018b}, where the bandwidth of individual sub-bursts in FRB 121012 is plotted against their centre frequency (ranging from $\sim 1$--$8$~GHz). The straight line is a linear fit to the data using the non-relativistic Doppler shift formula, resulting in a slope of $156\pm4$~MHz/GHz. \citet{Houde2018b} thus found that $\Delta\nu_\mathrm{obs}\simeq0.16\nu_\mathrm{obs}$, implying relativistic motions. However, this finding does not inform us on how strongly relativistic FRB 121102 is. It can be shown from equation (\ref{eq:Doppler}) that the relative frequency bandwidth measured in the observer's rest frame is linked to the velocity extent within the FRB source by
\begin{equation}
    \frac{\Delta\nu_\mathrm{obs}}{\nu_\mathrm{obs}} = \frac{\Delta\beta}{1-\beta^2}.\label{eq:bandwidth}
\end{equation}
This implies that even a small velocity extent $\Delta\beta$ can yield a significant observed bandwidth $\Delta\nu_\mathrm{obs}$, increasing with $\beta$, the velocity of the FRB source.

Although the scenario presented here is general in nature, it applies particularly well to the FRB model based on Dicke's superradiance developed in \citet{Houde2018a} and \citet{Houde2018b}. Superradiance is intrinsically a narrow-band process stemming from the coherent interaction between a gas hosting a population inversion and its radiation field \citep{Dicke1954,Gross1982,Rajabi2016A,Rajabi2016B,Rajabi2017,Rajabi2019,Rajabi2020}. For superradiance to happen the column density of the inverted gas must exceed a given threshold. This can proceed in different ways. For example, superradiance can be initiated through an increase in the intensity of the pumping source (usually at infrared wavelengths) that is responsible for the inversion, or by the action of a coherent incident field impinging on the region harbouring the inverted gas at the radiation frequency of the ensuing superradiance signal. Both types of excitation can act as triggers for the superradiance process, corresponding to the trigger source in our model (see Figure \ref{fig:cartoon}). It is important to note that the trigger can either be located externally, far from the FRB source or close by in its vicinity. The main requirement for our model is that the trigger is seen by the observer to be located behind the FRB source.

We also note that in their recent work \citet{Houde2018b} investigated the action of relatively weak incident signals from radio pulsars to trigger superradiant bursts. A continuum of behaviours are observed amongst pulsars and magnetars \citep{Camilo2006,Gavriil2008} which can act as a suitable trigger for initiating superradiance in line with their predictions. Overall, there is no discrimination against the precise nature of the trigger source, e.g., whether it consists of radio emission from magnetars or radio pulsars. This coherent enhancement of radiation via superradiance are suspected in a wide array of environments, including some potentially harbouring FRBs \citep{Houde2018a,Mathews2017,Rajabi2016A,Rajabi2016Thesis,Rajabi2020}. In particular, past surveys of magnetars suggest radio emissions are concomitant with increased X-ray activity of usually quiescent magnetars \citep{Camilo2006,Camilo2007}. With weak radio emissions from SGR~1935+2154 recently documented \citep{Kothes2018,Younes2017}, we suspect superradiance could be coherently amplifying radio pulses and be at the source of the recently detected bursts potentially emanating from this region of the Milky Way \citep{Scholz2020,CHIME2020b}. Within the context of our superradiance model, the apparent correlation between the X-ray peaks and FRB-like signals seen in this source (with the X-ray lagging the radio intensity bursts; \citealt{Mereghetti2020}) would imply a connection between the physics driving the X-ray emission and those driving the initial radio trigger.

Superradiance is characterized by a single time-scale $T_{\mathrm{R}}$, which sets the duration $\tau^\prime_\mathrm{w}$ of the superradiance pulse and $\tau^\prime_{\mathrm{D}}$ the time delay before its emergence, which is needed for coherence to be established in the radiating gas. This characteristic time-scale is proportional to the spontaneous emission time-scale of the underlying molecular or atomic transition, and varies inversely with the (square of the) wavelength of the radiation and the column density of the inverted population. In general,  $\tau^\prime_{\mathrm{D}}$ is expected to be approximately one and two orders of magnitude larger than the pulse duration and $T_{\mathrm{R}}$, respectively, in superradiant astronomical systems \citep{Rajabi2019,Rajabi2020}. This is consistent with the results of the fit shown in Figure \ref{fig:sub-burst_drift}, which implies $\tau^\prime_\mathrm{w}\simeq 0.1\tau^\prime_\mathrm{D}$.

It follows that the predictions made with the model discussed in this paper are contained within one based on Dicke's superradiance. In particular, the sad trombone requirement of $d\nu_\mathrm{obs}/d\tau^\prime_\mathrm{D}<0$ could be satisfied by systems where an increase in velocity is correlated with an increase in gas density (and the column density of the population inversion, resulting in lower $\tau^\prime_\mathrm{D}$ and $\tau^\prime_\mathrm{w}$; see above). An example could be a Keplerian accretion disk with increasing density at smaller radii. For such a scenario, superradiance systems would be observed along lines of sight parallel to well-defined tangential velocity fields (i.e., velocity components perpendicular to the plane of the sky (and the orbital radius) and thus directed toward the observer). Furthermore, the strong velocity coherence requirement for superradiance would imply that the radiation would be highly collimated in the same direction \citep{Rajabi2020}, even without taking into account radiation beaming from relativistic motions (which will further enhance this effect; see below). Such environments, which are similar to regions hosting megamasers \citep{Reid2009,Gray2012}, provide conditions consistent with our model where we limit our analysis to the case when the FRB source is moving in a direction (anti-)parallel to the line of sight from the observer. This is different from other relativistic models that consider situations involving smooth and continuous media with radiation emanating within comoving frames \citep{Beniamini2020}.  

Although this model would fit FRB 121012 and other repeating FRBs \citep{CHIME2019b,CHIME2019c,CHIME2020}, it does not have to be a rule for all FRBs and, therefore, will not preclude the observations of the opposite effect. As stated earlier, recent observations have revealed cases where a positive relative frequency drift is detected with successive sub-bursts (see \citealt{Scholz2020,CHIME2020b} for the detection in the direction of the Galactic magnetar SGR~1935+2154, and \citealt{Day2020} for FRB 190611).  
 
Relativistic motions must also be an integral component of a superradiance FRB model. For example, a source velocity of $\left|\beta\right|\lesssim0.8$ can produce signals of observed frequencies $0.3\nu_0\lesssim\nu_\mathrm{obs}\lesssim 3\nu_0$ and significant changes in bandwidths with $0.8\Delta\nu_0\lesssim\Delta\nu_\mathrm{obs}\lesssim 8\Delta\nu_0$ (using $\Delta\nu_0=\nu_0\Delta\beta$ as well as equations \ref{eq:Doppler} and \ref{eq:bandwidth}). Furthermore, and as previously noted and visualized in Figure 4 of \citet{Houde2018b}, the appearance of FRB spectra, although much broader in extent, is reminiscent of those observed in megamaser regions, which are also known to result from an underlying narrow-band emission process coupled to large systematic motions in their source. 

Advantageously, for a superradiance system the relativistic velocity of the FRB source enhances the probability of detection at cosmological distances, due to relativistic beaming and field amplification. In the beaming effect, the Lorentz transformation from the FRB rest frame to the observer frame reduces a given differential solid angle into which radiation propagates, centred upon the $\mathbf{e}_{x}$ direction in Figure \ref{fig:cartoon}, by a factor of $\left(1-\beta\right)/\left(1+\beta\right)$. In the field amplification effect, Lorentz transformation of the electric field (as a three-vector component of the electromagnetic field four-tensor) provides a further enhancement by a factor of $\left(1+\beta\right)/\left(1-\beta\right)$ for the flux density. The combination of both effects could bring an amplification by a factor as high as $\left[\left(1+\beta\right)/\left(1-\beta\right)\right]^{2}$ for sources radiating over a large enough solid angle. Evidently, the importance of this phenomenon increases as the velocity of the FRB source $\beta$ approaches unity. 

We also note that we did not consider the potential effects of scintillation and scattering on the relationships derived in our analysis. Although these phenomena can affect the temporal and spectral structures of FRB signals as they propagate from the sources to the observer, it is not clear, however, how they could alter the relationship between the sub-burst drift and duration, for example. The results obtained (e.g., see Figure \ref{fig:sub-burst_drift}) seem to imply that they do not for FRB 121102. Accordingly, we also note that it has been previously observed that some detected pulses for FRB 121102 did not display obvious signs of scintillation and scattering \citep{Scholz2016}.

Finally, while our model's primary objective was to explain the spectro-temporal structure common to FRB 121102 and other repeating FRBs, it is likely to have a broader reach in its applicability. That is, it will be instructive to verify if the law linking the sub-burst drift and duration given in equation (\ref{eq:sub_drift}) is widely observed in other FRBs. Such a potentially universal behaviour among FRBs would be extremely helpful to improve our understanding of the physical processes underlying this phenomenon, as well as being highly constraining for existing and future models.   

\section*{Acknowledgements}
M.H.'s research is funded through the Natural Sciences and Engineering Research Council of Canada Discovery Grant RGPIN-2016-04460. M.H. is grateful for the hospitality of Perimeter Institute where part of this work was carried out. F.R.'s research at Perimeter Institute is supported in part by the Government of Canada through the Department of Innovation, Science and Economic Development Canada and by the Province of Ontario through the Ministry of Economic Development, Job Creation and Trade. F.R. is in part financially supported by the Institute for Quantum Computing. M.A.C. is grateful to Victor Tranchant whose data loading and noise removal code was the starting point for the analysis pipeline. A.M. and C.M.W. are supported by the Natural Sciences and Engineering Research Council of Canada (NSERC) through the doctoral postgraduate scholarship (PGS D).

\section*{Software and Data}
The data pipeline is made available and maintained by M.A.C. at \url{https://github.com/mef51/sadtrombone}. Aggregate data of the bursts and the code for the figures are also available. Data of the FRB spectra are available either publicly or via the authors of their respective publications. The figures in this paper were prepared using the {\tt matplotlib} package \citep{Hunter2007}.

%%%%%%%%%%%%%%%%%%%%%%%%%%%%%%%%%%%%%%%%%%%%%%%%%%



\bibliographystyle{mnras}
\bibliography{SR-bib}

%%%%%%%%%%%%%%%%%%%%%%%%%%%%%%%%%%%%%%%%%%%%%%%%%%

%%%%%%%%%%%%%%%%% APPENDICES %%%%%%%%%%%%%%%%%%%%%

%\appendix


%%%%%%%%%%%%%%%%%%%%%%%%%%%%%%%%%%%%%%%%%%%%%%%%%%

% Don't change these lines
\bsp	% typesetting comment
\label{lastpage}
\end{document}